\documentclass[3p,10pt,fleqn]{elsarticle}
\usepackage{color}

\usepackage{graphicx}
\usepackage{bm,amsmath,amssymb}
\usepackage[mathscr]{eucal}
\usepackage{color}
\usepackage{afterpage}

\usepackage[figuresright]{rotating}
\long\def\comment#1{ }

\newcommand{\lan}{\langle}
\newcommand{\ran}{\rangle}
\newcommand{\abar}{\bar{\alpha}}

\newcommand{\eqn}[1]{Eq.~\eqref{#1}}
\newcommand{\beq}{\begin{equation}}
\newcommand{\eeq}{\end{equation}}
\newcommand{\nn}{\nonumber\\}

\newcommand{\rmd}{{\rm d}}
\newcommand{\rme}{{\rm e}}
\newcommand{\rmi}{{\rm i}}
\newcommand{\order}[1]{\mathcal{O}{(#1)}}
\newcommand{\rmI}{{\rm I}}

\newcommand{\bp}{\bm{p}}

\newcommand{\bx}{\bm{x}}
\newcommand{\by}{\bm{y}}

\newcommand{\bv}{\bm{v}}

\newcommand{\br}{\bm{r}}

\begin{document}

\begin{frontmatter}

\title{Jet quenching parameter in an expanding QCD plasma}
\author[addr1]{Edmond Iancu}
\author[addr1,addr2,addr3]{Pieter Taels}
\author[addr4]{Bin Wu}
%\email{b.wu@cern.ch}
\address[addr1]{Institut de Physique Th\'{e}orique, CEA Saclay, 91191, Gif-sur-Yvette Cedex, France}
\address[addr2]{
Department of Physics, University of Antwerp, Groenenborgerlaan 171, 2020 Antwerpen, Belgium}
\address[addr3]{INFN Sezione di Pavia, via Bassi 6, 27100 Pavia, Italy}
\address[addr4]{Theoretical Physics Department, CERN, CH-1211 Gen\`eve 23, Switzerland}

\begin{abstract}
We study the phenomenon of transverse momentum broadening for a high-$p_T$ parton propagating through a weakly-coupled quark-gluon plasma undergoing boost-invariant longitudinal expansion. We propose a boost-invariant description for this phenomenon, in which the broadening refers to the angular variables $\eta$ (the pseudo-rapidity) and $\phi$ (the azimuthal angle). The  jet quenching parameter $\hat{q}$ which enters this description depends upon the proper time alone. We furthermore consider radiative corrections to $\hat q$. As in the case of a static medium, we find potentially large corrections enhanced by a double logarithm. But unlike for the static medium, these corrections are now local in time: they depend upon the local (proper) time characterizing the expansion, and not upon the overall path length. We resum such corrections to all orders into a renormalized jet quenching parameter. The main effect of this resummation is to slow down the decrease of $\hat q$ with increasing proper time.

\end{abstract}

%% Preprint number  CERN-TH-2018-045

\end{frontmatter}
\vspace{0.5cm}

{\bf Introduction}.---The physics of ``jet quenching'', which globally denotes the modifications of the properties of an energetic jet or a hadron due to its interactions with a dense QCD medium, represents a main source of information about the quark-gluon plasma (QGP) produced in the intermediate stages of ultrarelativistic heavy ion collisions at RHIC and the LHC. This physics encompasses a large variety of physical phenomena, including transverse momentum broadening, energy loss via medium-induced radiation \cite{Baier:1996kr,Baier:1996sk,Baier:1998kq,Zakharov:1996fv,Wiedemann:2000za}, democratic branchings at large angles \cite{Blaizot:2013hx}, color decoherence of partonic radiators \cite{MehtarTani:2010ma,MehtarTani:2011tz,CasalderreySolana:2011rz}, or medium-induced constraints on the phase-space for vacuum-like emissions \cite{Caucal:2018dla}. Remarkably though, the theoretical descriptions of all these phenomena depend upon the properties of the medium via a {\it single} parameter, a transport coefficient known as the ``jet quenching parameter'' $\hat{q}$, which physically represents the rate for transverse momentum ($p_T$) broadening \cite{Baier:1996sk}. By itself, this  phenomenon of (medium-induced) $p_T$-broadening seems difficult to study in the data (at least, for sufficiently high energies), since it is hidden by the broadening introduced by vacuum-like, soft gluon radiation \cite{Mueller:2016gko,Mueller:2016xoc}. (The situation might be more favorable in the lower-energy environment at RHIC \cite{Mueller:2016gko,Chen:2016vem}; see e.g. a recent measurement by the STAR collaboration \cite{Adamczyk:2017yhe}.) However, the RHIC and LHC data for nuclear-nuclear collisions do show abundant evidence for the other medium effects alluded to above, notably for the radiative energy loss, via a multitude of observables like the nuclear modification factor for particle production \cite{Adcox:2001jp,Adler:2002xw,CMS:2012aa,Abelev:2012hxa}, the di-jet asymmetry \cite{Aad:2010bu,Chatrchyan:2011sx}, or the nuclear modification of the jet fragmentation function \cite{Chatrchyan:2014ava,Aad:2014wha}. Our physical understanding of these data depends in a crucial way upon our ability to provide reliable calculations for $\hat q$.

For a weakly coupled plasma --- the physical scenario that we shall assume in what follows (see also \cite{CasalderreySolana:2011us} for a review of calculations using the AdS/CFT correspondence at strong coupling and Refs.~\cite{Majumder:2012sh,Laine:2013lia,Panero:2013pla} for lattice approaches to $\hat q$) ---, the leading order contribution to the jet quenching parameter comes from elastic collisions with the plasma constituents \cite{Baier:1996sk,Arnold:2008vd}. The subleading effect of these collisions, which in a thermal plasma occurs at order $g$ due to the Bose enhancement of the soft thermal gluons, has been computed in \cite{CaronHuot:2008ni}.  The $\order{\alpha_s}$ radiative corrections to $\hat q$ have first been discussed in the exploratory studies \cite{CasalderreySolana:2007sw,Wu:2011kc}, followed by a seminal paper \cite{Liou:2013qya} which pointed out the existence of {\it double-logarithmic} corrections, of order $\alpha_s\ln^2(L/\lambda)$, with $L$ the distance travelled by the jet through the medium and $\lambda=1/T$ the wavelength of the thermal particles. Such corrections occur to all orders and can be reabsorbed into a redefinition of $\hat q$.  These conclusions have been comforted by subsequent studies \cite{Iancu:2014kga,Blaizot:2014bha, Wu:2014nca,Iancu:2014sha}, which also demonstrated the {\it universality} of the renormalization of $\hat q$ --- i.e. the fact that the same radiative corrections arise in the context of transverse momentum broadening \cite{Wu:2011kc,Liou:2013qya,Iancu:2014kga} and in that of the medium-induced radiation \cite{Iancu:2014kga,Blaizot:2014bha,Wu:2014nca}. Note however that this renormalization spoils the physical interpretation of $\hat q$ as a {\it local} transport coefficient: the renormalized $\hat q$ depends upon the overall path length $L$, due to the finite formation times of the quantum fluctuations, which can take any value between $\lambda$ and $L$.

All these recent studies of the radiative corrections to $\hat q$ have been performed for the case of a static medium. Clearly, this is not the actual physical situation in high-energy nucleus-nucleus collisions, where the partons liberated by the collision rapidly separate from each other along the collision axis --- the medium undergoes longitudinal expansion \cite{Bjorken:1982qr}. Moreover, all previous studies of the (collisional) transverse momentum broadening in such a longitudinally-expanding environment \cite{Baier:1998yf,Zakharov:1998wq,Arnold:2008iy,Wu:2010ze} were restricted, for simplicity,  to the case of hard probes (jets or partons) propagating at ``central rapidity'', i.e. perpendicular on the collision axis. This situation is somewhat simpler in that the only effect of the expansion is the time-dependence of $\hat q$: with increasing time, the plasma becomes more and more dilute, due to its expansion.

In this Letter, we shall consider a  longitudinally expanding plasma which is boost-invariant (a reasonably good approximation for the bulk matter created in heavy-ion collisions in the so-called ``central plateau'' region \cite{Bjorken:1982qr}, i.e. not very close to the collision axis) and generalize the previous studies in two respects. First, we shall reexamine the definition of the transverse momentum broadening in an expanding medium and provide a boost-invariant description for this phenomenon, which applies to hard probes propagating along arbitrary directions. In this more general description, the broadening refers to the angular variables $\eta$ (the pseudo-rapidity) and $\phi$ (the azimuthal angle), which are precisely the variables used to study the kinematics of the jets in practice. Furthermore, $\hat q$ enters as a rate for momentum\footnote{The momentum scale is provided by $p_T$ --- the component of the jet momentum which is transverse to the collision axis and which is not affected by the medium in the current approximations; see below for details.} broadening per unit {\it proper} time $\tau$; it decreases with increasing $\tau$, due to the medium expansion. Second, we compute the radiative corrections to $\hat q$ in this particular set-up and to double logarithmic accuracy. Interestingly, we find that the non-locality of the quantum corrections is {\em considerably reduced} by the expansion of the medium: the relevant quantum fluctuations are those whose formation time is at most of the order of the actual (proper) time $\tau$ of the collision. (Fluctuations with much larger formation times are unimportant due to the dilution of the medium.) Accordingly, the only effect of resumming these corrections is to modify the $\tau$-dependence of the jet quenching parameter: the renormalized $\hat q(\tau) $ is still {\it local} in (proper) time.

{\bf The longitudinally expanding plasma}.---We chose the $z$-axis as the collision (or ``longitudinal'') axis and use the notation $\bx_T=(x,y)$ for the coordinates of a point in the transverse plane. The collision starts at $z = t = 0$. We assume, as usual, that all the partons liberated by the collision are produced in a very short interval around $t=z=0$ and that the hydrodynamical expansion starts shortly after, at the ``thermalization'' time $\tau_0$. In a head-on collision of big nuclei, the transverse density of bulk matter does not variate significantly as long as it is not so far from the collision center (at a distance $x_T\lesssim R$ with $R$ the size of the nuclei). Accordingly, the transverse density can be taken to be homogeneous and the transverse components $v_x$ and $v_y$ of the fluid velocity can be taken to be zero. But these particles have non-trivial longitudinal momenta (or velocities), inherited from their parent nuclei, hence they will separate from each other along the $z$ axis and thus dilute the medium. We follow the standard assumption that this longitudinal expansion is {\em uniform}, thus leading to a  {\em boost-invariant particle distribution} \cite{Bjorken:1982qr} (see also \cite{Ollitrault:2008zz} for a pedagogical discussion):  all the particles that are found at $z$ at time $t$ have the same longitudinal velocity $v_z=z/t$, which is also the fluid velocity. Equivalently, the momentum rapidity\footnote{We treat all particles as massless, so their momentum rapidity is the same as their pseudo-rapidity: $\eta=-\ln\tan(\theta/2)$, with $\theta$ their polar angle: $v_z=\cos\theta$.} $\eta$ and the space-time rapidity $\eta_s$ of any of these particles, which in general are defined as
\begin{eqnarray}
\label{etadef}
\eta\equiv\frac{1}{2}\ln\left(\frac{E+p_z}{E-p_z}\right) ,\qquad \eta_s\equiv \frac{1}{2}\ln\left(\frac{t+z}{t-z}\right),
\end{eqnarray}
can be identified with each other, $\eta=\eta_s$, in the situation at hand. For such a longitudinally boost-invariant plasma, the parton density $\rho$ depends only upon the proper time $\tau\equiv\sqrt{t^2-z^2}$. As explained e.g. in Appendix A of Ref.~\cite{Baier:1998yf}, this $\tau$-dependence can be easily deduced for the case of an isentropic flow; one finds
\beq\label{Tt}
\left(\frac{T}{T_0}\right)^3\,=\,\left(\frac{\tau_0}{\tau}\right)^\beta,\qquad\mbox{with}\quad\beta\equiv 3v_s^2=\frac{1}{1+\Delta_1/3}\,.
\eeq
Here $v_s$ denotes the velocity of sound, which would be equal to $1/\sqrt{3}$ (implying $\beta=1$) for an ideal gas. The parameter $\Delta_1$, which measures the deviation from the ideal gas limit, is positive in perturbation theory and of order ${\alpha_s^2}$. That is, the power $\beta$ is close to, but strictly smaller than, one and the ideal gas limit $\beta\to 1$ will be considered too, for illustration.

{\bf The hard probe.}---Our test particle is an energetic parton (more generally, a jet) which is produced very early, near $t=z=0$, and also very centrally, at $x\simeq y\simeq 0$. It will be convenient to describe the kinematics of the on-shell partons by using its (pseudo)rapidity $\eta$ and its transverse momentum with respect to the beam axis, $\bp_T=(p_x,p_y) =p_T(\cos\phi, \,\sin\phi)$, with $\phi$ the azimuthal angle. From the definition \eqref{etadef} of $\eta$ together with the mass-shell condition $E= \sqrt{p_T^2+p_z^2}$, one easily deduces $E=p_T\cosh\eta$ and $p_z=p_T\sinh\eta$. Hence, we shall parametrize the initial 4-momentum of the parton as $t=0$ as 
\beq \label{p0}
p_0^\mu=p_T (\cosh\eta_0, \cos\phi_0,\sin\phi_0,\sinh\eta_0)\equiv p_T \hat{n}^\mu\,.
\eeq
At later times $\tau\gtrsim\tau_0$, when bulk matter gets formed, this parton suffers collisions off the medium constituents, with several physical consequences: a broadening  of its (transverse and longitudinal) momentum distribution, as well as (collisional and radiative) energy loss. Here we shall focus on the transverse momentum broadening (TMB), which is also the mechanism controlling energy loss via medium-induced radiation \cite{Baier:1996kr,Zakharov:1996fv,Wiedemann:2000za}. Notice that in the context of momentum broadening, the ``transverse plane'' is defined w.r.t. to the parton direction of motion, and not to the collision axis. We focus on a high-$p_T$ parton, whose momentum broadening due to multiple scattering is much smaller than its original $p_T$.

In the case of a {\em static} medium,  this broadening is naturally described as a random addition to the 3-momentum of the parton, which is orthogonal on its original direction of propagation in space (the same as its {\em average} direction in the presence of broadening). But in a longitudinally expanding medium, this definition becomes more subtle, because of the bias introduced by the fluid collective motion. The fact that the surrounding medium is invariant under a longitudinal boost makes it natural to use a boost-invariant picture for the TMB as well, that is, a picture in which the relevant time variable is the {\it proper} time $\tau$.

To construct such a picture, it is convenient to first consider a test particle which propagates at ``central rapidity'' ($\eta\simeq 0$, or $z\simeq 0$), i.e. in the plane transverse to the collision axis. In that case, $\tau=t$, so one can use the standard definition for TMB: if the original 4-momentum reads $p^\mu_0=p_{0T}(1,1,0,0)$, then its instantaneous version at some later time has additional, random, components in the $(y,z)$ plane, acquired via collisions in the medium: $p^\mu=(E,p_{x},p_y,p_z)$, with $p_x=p_{0T}$ and $|p_y|, |p_z|\ll p_{0T}$. In that case, the
``transverse momentum broadening''  refers to the distribution ${\rmd N}/{\rmd^2\bp_\perp}$ of the test particle with respect to the 2-dimensional vector $\bp_\perp\equiv (p_y,p_z)$. Both the transverse momentum $p_T=\sqrt{p_{0T}^2+p_y^2}$ and  the energy $E=\sqrt{p_{0T}^2+p_\perp^2}$ change only to second order, so one can identify $E\simeq p_T\simeq p_{0T}$ to the accuracy of interest. 

This discussion makes it natural to define the TMB for a parton with an arbitrary initial direction motion, cf. \eqn{p0}, as its broadening in $\eta$ and $\phi$ at (quasi)fixed $p_T$. This definition is indeed boost-invariant, since both the transverse momentum $p_T$ and the angular differences  $\Delta\phi\equiv \phi-\phi_0$ and $\Delta\eta\equiv \eta-\eta_0$ are invariant under a longitudinal boost. It is furthermore suitable for the phenomenology since in practice a jet is defined as a cluster of final-state particles with some distribution in the $\phi$-$\eta$ space. 

To summarize, the TMB to be considered in this paper is the dynamics which transforms the initial 4-momentum shown in \eqn{p0} into a 4-momentum with the following general form
\begin{align}\label{pmuboost}
p^\mu&=p_{T}\big(\!\cosh(\eta_0+\Delta\eta),\,\cos(\phi_0+\Delta\phi),\,\sin(\phi_0+\Delta\phi),\,\sinh(\eta_0+\Delta\eta)\big)\nn
&\simeq p_T \hat{n}^\mu- p_{T}\Delta\eta\,\hat\eta^\mu - p_{T}\Delta\phi\,\hat\phi^\mu\,,
\end{align}
where the expression in the second line holds to linear order in the small quantities $\Delta\eta$ and $\Delta\phi$. We have introduced here the two space-like 4-vectors
\begin{eqnarray}
\label{eq:phietaaxes}
\hat\eta^\mu\equiv (-\sinh\eta_0, 0, 0, -\cosh\eta_0),\qquad
\hat\phi^\mu\equiv (0,\sin\phi_0, -\cos\phi_0, 0),
\end{eqnarray}
which span the 2-dimensional vector space encoding the jet broadening in $\Delta\phi$ and $\Delta\eta$. These vectors are orthogonal in the 4-dimensional sense to the initial jet direction: $\hat\eta \cdot \hat{n} =\hat\phi\cdot \hat{n}=0$. For that reason, the respective components $p^\eta \equiv  \hat{\eta}\cdot p = p_{T}\Delta\eta$ and $p^\phi\equiv \hat{\phi}\cdot p = p_{T}\Delta\phi$ will be referred to as  ``transverse'' and collectively denoted with the subscript $\perp$ :  ${\bm p}_\perp\equiv(p^\phi,p^\eta)$. This 2-dimensional vector should not be confused with the {\it other} transverse momentum in the problem, namely $\bp_T=(p_x,p_y) $, which is orthogonal to the collision axis.

In the high-$p_T$ regime of interest, the multiple scattering responsible for TMB can be resummed to all orders within the {\it eikonal approximation}. To that aim, and in order to follow as closely as possible the respective calculations for the case of a static medium (see e.g. Sec. 4.1 in \cite{Iancu:2014kga}), it is convenient to use {\it light-cone coordinates} adapted to the (average) direction of motion of the test parton. Besides the light-like vector $\hat{n}^\mu$ and the two space-like 4-vectors $\hat\eta^\mu$ and $\hat\phi^\mu$ already introduced, we need another  light-like vector, conveniently chosen as
\begin{eqnarray}
\label{hattau}
{\hat{\tau}}^\mu\equiv \left(\cosh\eta_0, -\cos\phi_0, -\sin\phi_0,\sinh\eta_0\right),
\end{eqnarray}
and which obeys ${\hat{\tau}}\cdot \hat n=2$ and ${\hat{\tau}}\cdot \hat\eta={\hat{\tau}}\cdot \hat\phi=0$. For any 4-vector $v^\mu$, let us define its light-cone (LC) coordinates as $v^\mu=(v^+,v^-,\bv_\perp)$ with $v^+\equiv ({\hat{\tau}\cdot v})/{\sqrt{2}}$, $v^-\equiv ({\hat{n}\cdot v})/{\sqrt{2}}$, and of course $\bv_\perp=(v^\phi,v^\eta)$ with $v^\phi\equiv  \hat{\phi}\cdot v$ and $v^\eta \equiv  \hat{\eta}\cdot v$.  It is easy to check that that, in these new coordinates, the scalar product takes the expected LC form, that is, $v\cdot w=v^+w^- + v^- w^+ - v^\phi w^\phi-v^\eta w^\eta$. In particular, the LC form of the 4-momentum $p^\mu$, as inferred from the second line of \eqn{pmuboost}, reads $p^\mu=(\sqrt{2}p_T, 0,\bp_\perp)$, showing that the modulus $p_T$ of the transverse momentum plays the role of the LC longitudinal momentum in the new coordinates. Remarkably, this differs from the respective form of the {\it initial} 4-momentum, namely $p^\mu_0=(\sqrt{2}p_T, 0,{\bm 0}_\perp)$, only via the addition of the transverse component $\bp_\perp$ --- exactly as expected for TMB.

%\beq
%v^\mu=(v^+,v^-,v^\phi,v^\eta)\equiv \left(\frac{\hat{\tau}\cdot v}{\sqrt{2}},\frac{\hat{n}\cdot v}{\sqrt{2}},\hat{\phi}\cdot v,\hat{\eta}\cdot v\right),,\eeq

{\bf $\hat{q}$ as jet broadening in $\Delta\phi$ and $\Delta\eta$.}---We are now in a position to present the calculation of the transverse momentum distribution ${\rmd N}/{\rmd^2\bp_\perp}$ generated via soft multiple scattering for a test particle propagating along an arbitrary direction in a longitudinally expanding medium which is boost-invariant. For definiteness, we choose the test parton to be a quark, but the subsequent results immediately extend to a gluon after replacing the color factor $C_F=(N_c^2-1)/2N_c$ by $C_A=N_c$, with $N_c$ the number of colors.

As already mentioned, we shall employ the eikonal approximation, most conveniently formulated in transverse coordinate space (here, ``transverse'' = $\perp$), in which case the effect of multiple scattering is simply a rotation of the color state of the quark, as represented by a Wilson line in the fundamental representation:
\beq\label{WL}
V(\bx_\perp)\,=\,\exp\left\{ig\int_0^L \rmd\tau \,\hat{n}\cdot A_a(\tau\hat{n}+\bx_\perp)t^a\right\}\,=\,
\exp\left\{ig\int_0^{L^+} \rmd x^+ \, A^-_a(x^+, x^-=0,\bx_\perp)t^a\right\}.\eeq
We have written this Wilson line using both the original coordinates and the new LC coordinates. $A^-_a= (\hat{n}\cdot A_a)/\sqrt{2}$ is the projection of the color field representing the gluons in the medium along the direction of propagation of the test quark. In the absence of collisions, the energetic quark would follow the classical trajectory $x^\mu(\tau) = \tau\hat{n}^\mu$, or $x^+=\sqrt{2}\tau$, $x^-=0$ and $\bx_\perp=0$, with $0< \tau < L$ and $L^+=\sqrt{2}L$. ($L$ denotes the total proper time travelled by the quark through the medium.) The collisions lead to a broadening of the quark distribution in $\bp_\perp$; the respective amplitude can be computed by displacing the quark trajectory by a fixed amount $\bx_\perp$, as shown in \eqn{WL}, and then taking a Fourier transform $\bx_\perp\to \bp_\perp$. The respective cross-section is obtained by multiplying with a similar Wilson line in the complex conjugate amplitude, but at a different transverse coordinate $\by_\perp$, then tracing (averaging) the product of Wilson lines over the final (initial) color states, and finally taking the Fourier transform from $\br_\perp \equiv \bx_\perp-\by_\perp$ to $\bp_\perp$.  One thus finds
\begin{align}
 \label{ptbroad}
 \frac{\rmd N}{\rmd^2\bp_\perp} \,=\,
  \int \frac{\rmd^2\br_\perp}{(2\pi)^2}\,\rme^{-\rmi \bp_\perp \cdot \br_\perp} \,S(\br_\perp)\,,\quad
  \mbox{with}\quad S(\br_\perp)\equiv\,\frac{1}{N_c}\big\langle{\rm tr}\, V(\bx_\perp) V^\dagger(\by_\perp)\big\rangle,
 \end{align}
where the brackets in the definition of $S$ denote the averaging over the color fields in the medium, thus producing the plasma gluon distribution and its correlations. The quantity $S(\br_\perp)$ can be recognized as the $S$-matrix for the elastic scattering of a small quark-antiquark {\em color dipole} with transverse size ${\bm r}_\perp$.

For a weakly-coupled quark-gluon plasma, one can assume a Gaussian distribution for $A^-$, with long tails in $\bx_\perp$ which are cut off by Debye screening. On the other hand, the distribution in $x^+$ is quasi-local, due to Lorentz contraction (in the rest frame of the energetic probe). This is tantamount to saying that successive collisions can be treated as quasi-local and hence independent from each other. Then a standard calculation yields (see e.g. \cite{Iancu:2014kga})
 \begin{align}\label{Sdip1}
 {S}_0(\br_\perp)\,\simeq\,\exp\left\{-\, \pi\alpha_s^2 C_F 
 r_\perp^2\int_{\tau_0}^L\rmd \tau\,\rho(\tau) \ln\frac{1}{r_\perp^2m_D^2(\tau)}  \right\}\,,
  %\qquad
  % \hat q(1/r^2)\,\equiv\,4\pi\alpha_s^2 C_F n_0 \ln\frac{1}{r^2m_D^2}\,.
  \end{align}
where $\rho(\tau)$ denotes the density of the medium constituents and $m_D(\tau)$ the Debye screening mass; for a longitudinally expanding medium, both quantities depend upon the proper time alone. The logarithm $\ln(1/{r_\perp^2 m_D^2} )$ has been generated by the integration over the transverse momenta $q_\perp^2$ exchanged between the dipole and a medium constituent: this integral is logarithmic so long as $m_D^2 \ll q_\perp^2 \ll 1/r_\perp^2$. We implicitly assume here that $1/r_\perp\gg m_D$, as is indeed the case for the relevant values of $r_\perp$ (see below). 
 
The Fourier transform of ${S}_{0}(\br_\perp)$ in \eqn{Sdip1} is {\it a priori} complicated by the logarithmic dependence of the exponent upon $r_\perp^2$. This dependence is important if one considers an unusually hard scattering, which transfers a large transverse momentum $p_\perp\gg \mathcal{Q}_0(L)$. Here, $\mathcal{Q}_0^2(L)$ is the typical momentum squared acquired via multiple scattering, conveniently defined as the value of $1/r_\perp^2$ for which the exponent in \eqn{Sdip1}  becomes of order one; that is,
   \begin{align}\label{Q0}
   \mathcal{Q}_0^2(L)
   \,\equiv\,   4\pi\alpha_s^2 C_F \int_{\tau_0}^L\rmd \tau\,\rho(\tau) \ln\frac{\mathcal{Q}_0^2(L)}{m_D^2(\tau)}  
   \,\equiv\,  
   \int_{\tau_0}^L\rmd \tau\,\hat q_0 (\tau)
   \,.\end{align}
 But such rare collisions, leading to a large deflection, are not those that we generally associate with ``momentum broadening''. Rather, we are interested in the effect of many successive collisions, each of them transferring a relatively low momentum and collectively leading to a random walk in $\bp_\perp$. %
When $p_\perp\lesssim \mathcal{Q}_0(L)$, the integral in \eqn{ptbroad} is dominated by dipole sizes $r_\perp\sim 1/\mathcal{Q}_0(L)$ and can be evaluated by replacing $r_\perp \to 1/  \mathcal{Q}_0$ within the slowly-varying logarithm inside  \eqn{Sdip1}, which then becomes a Gaussian:
 \begin{align}\label{Sdiptree}
 {S}_{0}(\br_\perp)\,\simeq\,\exp\left\{-\frac{1}{4}\,\mathcal{Q}_0^2(L)\,\br_\perp^2\right\}.
  %\qquad
  % \hat q(1/r^2)\,\equiv\,4\pi\alpha_s^2 C_F n_0 \ln\frac{1}{r^2m_D^2}\,.
  \end{align}
Its Fourier transform is trivial and yields a Gaussian distribution in the transverse momentum $\bp_\perp$, or, equivalently (after using $\bp_\perp^2=p_T^2(\Delta\phi^2+\Delta\eta^2)$) in the angular deviations $\Delta\phi$ and $\rmd\Delta\eta$ :
\begin{align}
 \label{ptGauss}
 \frac{\rmd N}{\rmd^2\bp_\perp} \,\simeq\,\frac{1}{\pi \mathcal{Q}_0^2(L)}\,\rme^{-\frac{\bp_\perp^2}{\mathcal{Q}_0^2(L)}}
 \quad\Longleftrightarrow\quad \frac{\rmd N}{\rmd\Delta\phi\, \rmd\Delta\eta} \,\simeq\,\frac{p_T^2}{\pi \mathcal{Q}_0^2(L)}\,\rme^{-\frac{p_T^2(\Delta\phi^2+\Delta\eta^2)}{\mathcal{Q}_0^2(L)}}
\, .\end{align}
 
As shown by the second equality in \eqn{Q0}, it is natural to interpret the integrand there as a 
(quasi) local transport coefficient, the {\it jet quenching parameter}  $\hat q_0$, which  represents the average transverse momentum squared transferred per unit {\it proper} time.  (We use a subscript 0 on $\hat q_0$ to remind that this quantity refers to the tree-level approximation, to be later amended by radiative corrections.) In the semi-classical approximation at hand, this transport coefficient is proportional to the rate for large-angle scattering. As visible in \eqn{Q0}, $\hat q_0$ is not fully local, but rather  {\em quasi}-local: it has a weak, logarithmic, dependence upon the global distance $L$ (that will be kept implicit in our notations), due to the non-locality of the Coulomb exchange. For a static plasma with density $\rho_0$, one finds $\mathcal{Q}_0^2(L)=\hat q_0 L$ with
 \begin{align}\label{Q0static}
 \hat q_0 \,=\,   4\pi\alpha_s^2 C_F \rho_0\ln\frac{\mathcal{Q}_0^2(L)}{m_D^2} 
   \,.\end{align}
For the more interesting case of a longitudinally expanding plasma, whose density $\rho(\tau) \propto T^3(\tau)$ is decreasing in time according to \eqn{Tt}, we can neglect the additional time-dependence associated with the lower limit $m_D^2(\tau)$ of the logarithm in \eqn{Q0}.  Under this approximation, one can write
\beq\label{qt}
\hat q_0 (\tau)\,\simeq\,\hat q_0 (\tau_0)\left(\frac{T(\tau)}{T_0}\right)^3\,\simeq\,\hat q_0(\tau_0)\left(\frac{\tau_0}{\tau}\right)^\beta.\eeq
Then the integral in \eqn{Q0} can be explicitly performed, to yield
\beq\label{Qst}
\mathcal{Q}_0^2(L)\,=\,\int_{\tau_0}^L\rmd\tau\,\hat q_0 (\tau)\,\simeq\,\hat q_0 (\tau_0) \tau_0^\beta\,\frac{L^{1-\beta}-\tau_0^{1-\beta}}
{1-\beta}\,\simeq\,\hat q_0 (L)L\,\frac{1-(\tau_0/L)^{1-\beta}}
{1-\beta}\,,
\eeq
where in the last equality we have used again \eqn{qt} for $\hat q_0(\tau)$. It is interesting to consider two limits of this result: \texttt{(i)} the large time limit $L\gg \tau_0$ for $\beta< 1$, and \texttt{(ii)} the ideal gas limit $\beta\to 1$. One finds
\beq
\label{Q0L}
   \mathcal{Q}_0^2(L)\simeq \begin{cases}
        \displaystyle{
        \frac{\hat q_0(L) L}{1-\beta}} & \mbox{for $\beta< 1$\,,}
        \\*[0.2cm]
        \displaystyle{\hat q_0(\tau_0)\tau_0\,\ln\frac{L}{\tau_0}} &
       \mbox{for $\beta=1$\,,}
    \end{cases}
    \eeq
where we have also used the fact that the quantity $\hat q (\tau)\tau$ is independent of $\tau$ when $\beta=1$. In both cases, one sees that the growth of the saturation momentum with $L$ is significantly slowed down by the expansion (and the ensuing dilution) of the medium.

{\bf Radiative corrections to $\hat{q}$.}---We now turn to the second issue that we would like to address in this paper, namely the calculation of the (dominant) radiative corrections to the jet broadening in $\Delta\phi$ and $\Delta\eta$ for the case of a medium undergoing (boost-invariant) longitudinal expansion. As we shall see, our main conclusion and also most of the details of the calculation are quite similar to those corresponding to a static medium \cite{Wu:2011kc,Liou:2013qya,Iancu:2014kga,Blaizot:2014bha} . Notably, we shall find a class of radiative corrections which are large since enhanced by  double (energy and collinear) logarithms, and which can be resummed to all orders into a renormalization of the jet quenching parameter. But we shall also find some noticeable differences, coming from the fact that the scales which control the phase-space for quantum corrections are now changing with time, due to the expansion. Most interestingly, we shall see that the dominant radiative corrections look {\em quasi-local} on the time scale for the expansion.

Physically, the radiative corrections to the transverse momentum broadening express the recoil accumulated by the incoming parton (here, a quark) via medium-induced gluon emissions. Clearly, the largest such contributions come from emissions which are relatively hard, in the sense of having a large transverse momentum $k_\perp$ w.r.t. the ``jet axis'' (the direction of motion of the original quark). The logarithmic enhancement can be roughly understood as follows (see  \cite{Wu:2011kc,Liou:2013qya} for more details): at high $k_\perp$, the medium-induced radiation is controlled by a {\it single,  relatively hard, scattering}, leading to a power-law tail $\propto 1/k_\perp^4$ in the gluon spectrum ${\rmd N}_g/({\rmd\omega\rmd k_\perp^2})$, 
together with the usual, bremsstrahlung, enhancement $\propto 1/\omega$ at low energies. Accordingly, the recoil contribution to the $p_\perp$-broadening of the parent quark, estimated as
 \beq\label{deltap|}
 \delta\lan p_\perp^2\ran =\int \frac{\rmd\omega}{\omega}
 \int {\rmd k_\perp^2}\,{k_\perp^2}\,\frac{{\rmd N}_g}{\rmd\omega\rmd k_\perp^2}
 \eeq
is enhanced by two potentially large logarithms: an energy logarithm and a transverse momentum one. The importance of this enhancement depends upon the precise arguments of the two logarithms, which are in turn determined by the phase-space for gluon radiation triggered by a single scattering. In what follows, we shall motivate this phase-space via physical considerations, which are ultimately supported by the explicit calculations presented in the Appendix. For more clarity, we shall first consider the case of a static medium. 

For an emission to be insensitive to multiple scattering, the gluon momentum $k_\perp$ should be much larger than the momentum that could be acquired via collisions in the plasma during the ``formation time'' $t_{\rm f}\simeq 2\omega/k_\perp^2$  (the typical duration of the emission): $k_\perp^2\gg \hat q_0 t_{\rm f}$. On the other hand, $k_\perp$ should be much smaller than the overall  transverse momentum broadening $\lan p_\perp^2\ran \simeq \hat q_0 L$ of the quark, since the recoil represents only a small contribution to the latter. Hence, the transverse logarithm in the case of a static medium comes from transverse momenta within the range $ \hat q_0 t_{\rm f} \ll k_\perp^2\ll \hat q_0 L$. This transverse phase-space shrinks to zero when $t_{\rm f}$ becomes comparable with $L$.  This condition can be used to deduce an upper limit on the gluon energy $\omega$, but as a matter of facts it is preferable to use the formation time  $t_{\rm f}$ itself as the variable characterizing the longitudinal phase-space. Specifically, the temporal (or energy) logarithm comes from the range $\lambda\ll t_{\rm f} \ll L$, where the lower limit, which features the thermal wavelength $\lambda=1/T$ of a typical medium constituent, can be understood as follows: for the gluon to suffer a collision during its emission, the formation time should be at least as large as the longitudinal extent $\sim \lambda$ of a parton from the medium \cite{Liou:2013qya,Iancu:2014kga}.

These considerations motivate the following result for the one-loop correction to the transverse momentum broadening in a static medium and in the double-logarithmic approximation (DLA) \cite{Liou:2013qya,Iancu:2014kga,Blaizot:2014bha}:
    \begin{align}
\label{DLAstatic}
\delta\lan p_\perp^2(L)\ran 
\,=\,% \frac{2\alpha_s C_F}{\pi} 
\abar\hat q_0\int_0^L\rmd t \int_{\lambda}^{L}
\frac{\rmd t_{\rm f}}{t_{\rm f}}
  \int^{\hat q_0 L}_{\hat q_0t_{\rm f}} \frac{\rmd k_\perp^2}{k_\perp^2} 
  \,=\, \hat q_0 L\,
\frac{\abar}{2}
 \ln^2\frac{L}{\lambda} 
  \,\equiv\,
  \delta \hat q(L) L\,,
  \end{align}
where $\abar\equiv{\alpha_s N_c}/{\pi}$ is the QCD coupling at the emission vertex and the third integral runs over the time $t$ at which the fluctuation suffers a collision in the medium: this time can take any value between 0 and $L$ and the gluon emission is localized within an interval $t_{\rm f}$ around it. The overall result for $\delta\lan p_\perp^2(L)\ran$ is roughly linear in $L$, so it can be also interpreted as a change $\delta \hat q(L)$ in the jet quenching parameter, as shown in the last equality in \eqn{DLAstatic}. This change is strictly speaking non-local, due to the fact that gluon emissions require a finite formation time, but this non-locality is very weak ($\delta \hat q(L)$ is only logarithmically sensitive to the path length $L$), because the typical values of $t_{\rm f}$ which matter at DLA are much smaller than $L$.

Turning now to a longitudinally-expanding medium and for a test quark propagating along an arbitrary direction, cf.  \eqn{p0}, it is useful to notice that the calculation resembles very much the one for a static medium provided one uses the ``right'' kinematical variables, that is, the LC coordinates introduced below \eqn{hattau} and which are better suited for a boost-invariant set-up. Specifically, the ordinary time variable $t$ in the static medium case should be replaced\footnote{To shed more light on this correspondence, one can observe that, already for a static medium, the time variable $t$ is essentially the same as the LC time, $x^+\simeq \sqrt{2}t$, and similarly $k^+\simeq \sqrt{2}\omega$, with the LC variables defined in the standard way for a leading parton with initial 4-momentum $p^\mu_0=p(1, \vec v)$; that is, $x^\pm\equiv (t\pm\vec v \cdot\vec x)/\sqrt{2}$. For the expanding medium and with the new LC coordinates defined below \eqn{hattau}, one rather has $x^+\simeq \sqrt{2}\tau$, which makes it natural to identify the variables $t$ and $\tau$ in the two problems.} by the proper time $\tau$ (and similarly $t_{\rm f}\to \tau_{\rm f}$) and the transverse momenta like $p_\perp$ and $k_\perp$ should be understood in the sense of \eqn{eq:phietaaxes}, e.g. ${\bm k}_\perp\equiv(k^\phi,k^\eta)$. This correspondence is also supported by the explicit calculations in the Appendix.

As before, the transverse logarithm is generated by integrating over gluon transverse momenta $k_\perp^2$ which are much larger than the momentum broadening $\Delta k_\perp^2(\tau_{\rm f})$ that would be accumulated during the formation time, but much smaller than the overall momentum broadening $\lan p_\perp^2(L)\ran \simeq \mathcal{Q}_0^2(L)$ accumulated by the leading quark over a (proper) time $L$. The upper limit $\mathcal{Q}_0^2(L)$ has already been computed (in the semi-classical approximation) in \eqn{Q0L}, so let us concentrate on the lower limit. This can be estimated as
\beq\label{kf}
\Delta k_\perp^2(\tau_{\rm f}) = \int_\tau^{\tau+\tau_{\rm f}}\rmd \tau'\,\hat q_0(\tau')\,\simeq\,\hat q_0(\tau)\tau_{\rm f}\,,\eeq
where we considered an emission which is initiated at $\tau$ and completed at $\tau+\tau_{\rm f}$. In writing the last estimate, we have anticipated that to the double-logarithmic accuracy of interest, the formation time $\tau_{\rm f}$ of a typical emission is much smaller than the absolute time $\tau$ when the fluctuation is initiated. It is furthermore instructive to separately consider the transverse logarithm, first for the case where the medium is an ideal QGP (i.e., $\beta=1$  in \eqn{Q0L})
\beq\label{intkT}
 \int^{\mathcal{Q}_0^2(L)}_{\hat q_0(\tau)\tau_{\text{f}}} \frac{\rmd k_\perp^2}{k_\perp^2}\,=\,
 \ln\frac{\mathcal{Q}_0^2(L)}{\hat q_0(\tau)\tau_{\text{f}}} \,\simeq\,\ln\frac{\tau}{\tau_{\text{f}}}+
 \ln\ln\frac{L}{\tau_0}\,\simeq\,\ln\frac{\tau}{\tau_{\text{f}}}\quad\mbox{for}\quad\beta=1\,.
 \eeq
In the last estimate above we kept only the term which will contribute to DLA. 
Notice that this term is independent of $L$, due to the fact that the $L$-dependence of the upper limit $\mathcal{Q}_0^2(L)$ is very weak for an expanding plasma. Clearly, this last estime makes sense if and only if $\tau$ is sufficiently large compared to $\tau_{\rm f}$ (a property that we have anticipated in relation with \eqn{kf}). That is, a  transverse logarithm can occur only for fluctuations which are emitted promptly enough on the typical time scale for the medium expansion (the same as the absolute time $\tau$). Fluctuations with much larger formations times are irrelevant (to the accuracy of interest), due to the medium dilution through expansion.

These considerations explain why the proper generalization of \eqn{DLAstatic} to the case of an expanding medium reads
  \begin{align}
\label{DLAexp}
\delta\lan p_\perp^2(L)\ran 
\,=\,% \frac{2\alpha_s C_F}{\pi} 
\abar\int_{\tau_0}^L \rmd \tau\,\hat q_0(\tau) \int_{\lambda(\tau)}^{\tau}
\frac{\rmd \tau_{\rm f}}{\tau_{\rm f}}
  \int^{\mathcal{Q}_0^2(L)}_{\hat q_0(\tau)\tau_{\rm f}} \frac{\rmd k_\perp^2}{k_\perp^2} \,.
  \end{align}
Strictly speaking, this expression has been here justified for the limiting case $\beta=1$, but it remains valid when $\beta < 1$, since in that case the final integral over $\tau$ is controlled by its upper limit $\tau=L$, hence there is no significant difference between choosing $\tau$ or $L$ as the upper limit in the integral over $\tau_{\rm f}$. In particular, it can be easily checked that, to DLA, the transverse integral is again given by the last estimate in \eqn{intkT}. Using this estimate, it is easy to see that the one-loop correction to the transverse momentum broadening for the expanding medium is consistent with a {\it local} renormalization $\delta\hat q(\tau) $ of the jet quenching parameter:
 \begin{align}
\label{hatqexp}
\delta\lan p_\perp^2(L)\ran 
\,=\,% \frac{2\alpha_s C_F}{\pi} 
\int_{\tau_0}^L\rmd \tau\,\delta\hat q(\tau),\qquad\mbox{with}\qquad
\delta\hat q(\tau)\,\equiv\,\hat q_0(\tau) \,\frac{\abar}{2} \ln^2\frac{\tau}{\lambda(\tau)}\,.
  \end{align}
The final result for $\delta\lan p_\perp^2(L)\ran$ depends upon $\beta$, because of the respective dependence of the tree-level piece $\hat q_0(\tau)$, cf. \eqn{qt} , and of the thermal wavelength 
$\lambda(\tau)=\lambda_0(\tau/\tau_0)^{\beta/3}$, with $\lambda_0\equiv 1/T_0$. One finds
\beq
\label{deltaQexp}
\delta\lan p_\perp^2(L)\ran\simeq \begin{cases}
         \displaystyle{\mathcal{Q}_0^2(L)\,\frac
{2\abar}{27}\,\ln^2\frac{L}{\tau_0}} &
       \mbox{for $\beta=1$\,,}
        \\*[0.2cm]
          \displaystyle{
        \mathcal{Q}_0^2(L)\,\frac
{\abar}{2}\left(1-\frac{\beta}{3}\right)^2\ln^2\frac{L}{\tau_0}}& \mbox{for $\beta< 1$\,,}
    \end{cases}
    \eeq
with $\mathcal{Q}_0^2(L)$ given by \eqn{Q0L} and  $L\gg \tau_0$.
In obtaining these results, we have neglected the difference between $\tau_0$ and $\lambda_0$ within the arguments of the various logarithms, to simplify writing. 

\eqn{deltaQexp} presents the dominant quantum correction to the characteristic scale $\mathcal{Q}_0^2(L)$ within the $\bp_\perp$-distribution in \eqn{ptGauss}. For instance, when $\beta < 1$, one can define a renormalized version of that scale, including the above radiative correction, as follows:
\beq\label{renQs}
\mathcal{Q}^2(L)\equiv \mathcal{Q}_0^2(L)\left\{1+\frac{\abar}{2}\left(1-\frac{\beta}{3}\right)^2\ln^2\frac{L}{\tau_0}\right\}.
\eeq
But for applications to other physical problems, like the medium-induced radiation, it is useful to also  have a local (in time) version of this renormalization, that is, a renormalized jet quenching parameter; for any $\beta\le 1$, this reads  (cf. \eqn{hatqexp})
  \beq\label{renqhat}
 \hat q(t) \equiv \hat q_0(\tau)+ \delta\hat q^{(1)}(\tau) = \hat q_0(\tau)\left\{1+\frac{\abar}{2}\ln^2\frac{\tau}{\lambda(\tau)}\right\}\,,\eeq
where the upper script ``$(1)$'' on  $\delta\hat q^{(1)}$ is meant to emphasize that this is a one-loop correction.
 
The experience with perturbative QCD teaches us that the appearance of an one-loop correction enhanced by a double (collinear and energy) logarithm signals the existence of a tower of higher-order such corrections --- here, $n$-loop corrections of order $\big(\abar\ln^2(\tau/\lambda)\big)^n$  ---, which need to be resummed whenever the double logarithm is large, $\abar\ln^2(\tau/\lambda)\gtrsim 1$. Such corrections, that were explicitly computed and resummed for the case of a static medium \cite{Liou:2013qya,Iancu:2014kga,Blaizot:2014bha,Iancu:2014sha}, are generated by successive gluon emissions which are strongly ordered in both formation times and transverse momenta: $\tau_{\text{f}}$ and $k_{\perp}^2$ are both strongly decreasing from one emission to the next one. By following the same steps as in Refs. \cite{Liou:2013qya,Iancu:2014kga,Iancu:2014sha}, it is straightforward to carry out the all-order resummation of the leading double-logarithmic terms.  One thus  obtains
 \beq
 \label{qyy}
 \hat{q}(\tau) = \hat{q}_0(\tau)\,
 \frac{\rmI_1\big(2 \sqrt{\abar}\, Y\big)}{\sqrt{\abar}\,Y}
 = \hat{q}_0(\tau)\, 
 \frac{\rme^{2 \sqrt{\abar}\, Y}}{\sqrt{4\pi}\,(\sqrt{\abar}Y)^{3/2}}
 \left[1 + \mathcal{O}(1/\sqrt{\abar}Y) \right],
 \eeq    
where $Y=\ln(\tau/\lambda(\tau))$, $\rmI_1$ is the modified Bessel function of the first kind, and the second equality holds when $Y \gg 1/\sqrt{\abar}$, i.e. for sufficiently large time. As manifest on the above equation, the main effect of the resummation is to slow down the decrease of $\hat q(\tau)$ with $\tau$. For instance, in the large time regime where the asymptotic expansion applies, the effective power for this decrease is reduced from $\beta$ down to $\beta- 2 \sqrt{\abar}\big(1-\beta/3\big)$. In particular, for an ideal plasma, this amounts to $1\to 1-4 \sqrt{\abar}/3$. Clearly, this  represents a significant reduction for a realistic value of the coupling like $\abar=0.3$.

\eqn{qyy} is formally similar to the corresponding result for a static medium  \cite{Liou:2013qya,Iancu:2014kga,Blaizot:2014bha,Iancu:2014sha}. It differs from the latter merely via the $\tau$-dependence of the functions $\hat{q}_0(\tau)$ and $\lambda(\tau)$ and, especially, via the replacement of $L$ (the global distance traveled through the medium by the leading particle) by $\tau$ (the time of scattering, as measured from the beginning of the expansion) as the upper time scale in the argument of the logarithm. This last replacement makes it possible to treat the renormalized $\hat q(\tau)$ as a {\em (quasi)local} transport coefficient, like its tree-level counterpart $\hat{q}_0(\tau)$.  
 
 \bigskip
\noindent{\bf Acknowledgements} We are grateful to Al Mueller for a critical reading of the manuscript and many useful discussions. P.T. and B.W. would like to thank Institut de Physique Th\'eorique de Saclay for hospitality during the gestation of this work. The work of E.I. is supported in part by the Agence Nationale de la Recherche project ANR-16-CE31-0019-01.   P.T. is funded by the European Union's Horizon 2020 research and innovation programme (grant agreement No. 647981, 3DSPIN).

\appendix

\section{One-gluon correction to momentum broadening in the expanding plasma}

In this section, we shall succinctly present the calculation of the one-loop (one gluon emission) radiative correction to the transverse momentum broadening in the expanding plasma, with the purpose of providing an explicit mathematical proof for \eqn{DLAexp}. Our calculation will closely follow the corresponding calculation for the case of a static medium in Refs.~\cite{Wu:2011kc,Liou:2013qya}, as well as the original calculation of medium-induced radiation in the expanding medium  \cite{Baier:1998yf}. We shall use the dipole picture together with the light-cone coordinates  introduced below \eqn{hattau}, in which $x^+$ is essentially the proper time ($x^+\simeq \sqrt{2}\tau$). Within this picture, the one-loop radiative correction corresponds to Feynman graphs in which a soft gluon, with longitudinal momentum $k^+\equiv \omega$ much smaller than the corresponding momentum of the incoming quark, is emitted by the dipole at some (proper) time $\tau_1$ and then reabsorbed at $\tau_2$, with $0\le \tau_1 < \tau_2 \le L$. The respective contribution to the momentum broadening of the quark, cf. \eqn{deltap|}, can be obtained as
 \beq
 \label{deltap}
\delta\lan p_\perp^2\ran = -\nabla_{\br}^2 \delta S(\br)\Big |_{r_\perp=0}\,,\eeq
where $ \delta S(\br)$ (with $\br\equiv \br_\perp$) represents the one-loop correction to the dipole $S$-matrix, which arises from the fact that  the system which scatters at intermediate times $\tau_1 <\tau <\tau_2$ is not a dipole anymore, but a system of 3 partons: a quark, an antiquark, and a gluon. The essential ingredient of the calculation is therefore the Green function $G^{(3)}( {\bm B}_{2}, \tau_2; {\bm B}_{1}, \tau_1;\br)$ describing the diffusion of this 3-parton system in the transverse space (${\bm B}_{2}$ and ${\bm B}_{1}$ denote generic transverse coordinates, in the sense of \eqn{eq:phietaaxes}). This is the solution to the following equation (in the large $N_c$ limit) \cite{Wu:2011kc,Liou:2013qya}
\begin{align}
i \frac{\partial}{\partial \tau} G^{(3)}( {\bm B}, \tau; {\bm B}_{1}, \tau_1;\br)
= \left\{\frac{1}{2 \omega}\nabla_{\bm B}^2- \frac{i\hat{q}_0(\tau)}{4} \left[ {\bm B}^2 + ({\bm B} - \br)^2 \right]\right\}G^{(3)}( {\bm B}, \tau; {\bm B}_{1}, \tau_1;\br),\label{equ:evotionoff}
\end{align}
with the initial condition $G^{(3)}( {\bm B}_{2}, \tau_1; {\bm B}_{1}, \tau_1) =\delta^{(2)}( {\bm B}_{2}-{\bm B}_{1})$. \eqn{equ:evotionoff} is related to the Schr\"odinger equation for a two-dimensional harmonic oscillator with an imaginary potential. Its solution can be constructed as in Appendix B of \cite{Baier:1998yf} and reads $G^{(3)}( {\bm B}_{2}, \tau_2; {\bm B}_{1}, \tau_1;\br)=G\left({\bm B}_2 - \frac{\bm r}{2},\tau_2;{\bm B}_1 - \frac{\bm r}{2},\tau_1\right)$,
with
\begin{eqnarray}
G({\bm B}_2, \tau_2; {\bm B}_1, \tau_1)\equiv \frac{i \omega}{2\pi D(\tau_2,\tau_1)}\,\rme^{\frac{-i\omega}{2 D(\tau_2, \tau_1)}[c_1 {\bm B_1^2}+c_2 {\bm B_2^2}-2 {\bm B}_2\cdot {\bm B}_1]},
\end{eqnarray}
with $c_1\equiv c(\tau_2,\tau_1), c_2\equiv c(\tau_1,\tau_2)$ and the following definitions for the functions $D(\tau_2,\tau_1)$ and $c(\tau_2,\tau_1)$ :
\begin{align}\label{Dcdef}
D(\tau_2,\tau_1)&=\pi  \nu  \sqrt{\tau _1 \tau _2}\, \big[J_{\nu }\left(2 \nu  \Omega_1 \tau _1\right) Y_{\nu }\left(2 \nu \Omega_2 \tau _2\right)-J_{\nu }\left(2 \nu  \Omega_2 \tau _2\right) Y_{\nu }\left(2 \nu  \Omega_1 \tau _1\right)\big],\nonumber\\
c(\tau_2,\tau_1)&=\frac{\pi  \nu  \sqrt{\tau _1 \tau _2}\Omega_2}{\sin(\pi  \nu )}\,\big[J_{\nu -1}\left(2 \nu  \Omega_2 \tau _2\right) J_{-\nu }\left(2 \nu  \Omega _1 \tau _1\right)+J_{1-\nu }\left(2 \nu \Omega _2 \tau _2\right) J_{\nu }\left(2 \nu\Omega_1\tau _1\right)\big],
\end{align}
where $J_{\nu }$, $Y_{\nu }$ etc. are the standard Bessel functions and we have used the shorthand notation $\Omega_{1,2}\equiv \Omega(\tau_{1,2})$ with $\Omega(\tau)\equiv \sqrt{i\hat{q}_0(\tau)/\omega}$ and $\nu\equiv 1/(2-\beta)$. For $\beta=0$ (hence $\nu=1/2$), these equations reduce to the expected result for the case of a static medium, that is, Eq.~(8) in \cite{Liou:2013qya}.

Given the 3-body Green function, the one-loop correction $ \delta S(\br)$ to the dipole amplitude is computed as 
\begin{align}
\delta S({\bm r})&= - \frac{\alpha_s N_c r^2}{2}
~\text{Re}~\int \frac{\rmd\omega}{ \omega^3}  \int_{\tau_0}^{L} \rmd\tau_2 \int_{\tau_0}^{\tau_2} \rmd\tau_1\nonumber\\
&\times \left\{
\rme^{-\frac{{\bm r}^2}{4}\int_{\tau_0}^{L} \rmd\tau'\hat{q}_0(\tau')[\theta(\tau_1-\tau')+\theta(\tau'-\tau_2)] } \nabla_{{\bm B}_2} \cdot \nabla_{{\bm B}_1} G^{(3)}( {\bm B}_{2}, \tau_2; {\bm B}_{1}, \tau_1;\br)%\right.\nonumber\\
%&-&\left.\left. \nabla_{{\bm B}_2} \cdot \nabla_{{\bm B}_1}  G_0( {\bm B}_{2}, t_2; {\bm B}_{1}, t_1 )
\right\}\Bigg |^{{\bm B}_2 = {\bm r}}_{{\bm B}_2 = 0}\,
\Bigg |^{{\bm B}_1 = {\bm r}}_{{\bm B}_1 = 0},\label{equ:Sfull}
\end{align}
where it is understood that one has to subtract the vacuum limit $\hat q_0\to 0$ of the integrand. To evaluate the correction \eqref{deltap} to the momentum broadening, one can expand \eqn{equ:Sfull} to lowest order in $\br^2$, to deduce\footnote{Strictly speaking, the small-$r^2$ expansion in \eqn{equ:Sexp} introduces a logarithmic ultraviolet divergence, but to the accuracy of interest, this can be simply dealt with by limiting the energy integration to $\omega <
(\tau_{\text{f}}/r^2)\simeq \tau_{\text{f}}\mathcal{Q}_0^2(L)$.}
\begin{align}
\delta S({\bm r})&\simeq - \frac{\alpha_s N_c r^2}{4}
~\text{Re}~\int \frac{\rmd\omega}{ \omega^3}  \int_{\tau_0}^{L} \rmd\tau_2 \int_{\tau_0}^{\tau_2} \rmd\tau_1\nonumber\\
&\qquad \quad \times \big( \nabla_{{\bm B}_2} \cdot \nabla_{{\bm B}_1}\big)^2 
\big[ G( {\bm B}_{2}, \tau_2; {\bm B}_{1}, \tau_1)- G_0( {\bm B}_{2}, \tau_2; {\bm B}_{1}, \tau_1 )\big]\bigg |_{{\bm B}_1= {\bm B}_2 = 0},\label{equ:Sexp}
\end{align}
where the vacuum subtraction, which involves the free propagator $G_0$, is now explicit. 

To extract the double-logarithmic contribution to \eqn{equ:Sexp}, it is convenient to change the time variables according to $\tau\equiv (\tau_1+\tau_2)/2$ and $\tau_{\text{f}}\equiv \tau_2-\tau_1$, implying $0 <\tau_{\text{f}} < \tau < L$. Specifically, the temporal logarithm arises from integrating over relatively small formation times $\tau_{\text{f}} \ll \tau$. In this limit, the functions $D$ and $c$ in \eqn{Dcdef} are formally the same as for the static medium, but with a time-dependent $\hat q$ :
%(evaluated at the average emission time $\tau$):% i.e. $\hat q(\tau)$:
\begin{eqnarray}\label{Dcapp}
D(\tau_2,\tau_1)\simeq \frac{\sin \left( \Omega(\tau) \tau_{\text{f}}\right)}{\Omega(\tau)},\qquad c(\tau_2,\tau_1) \simeq \cos \left(\Omega(\tau)  \tau _{\text{f}}\right).
\end{eqnarray}
Accordingly, the subsequent analysis of \eqn{equ:Sexp} involves exactly the same manipulations as in the case of the static medium,  except for the replacement  $\hat q_0\to \hat q_0(\tau)$ and for the fact that the integral over the formation time should now be restricted to  $\tau_{\text{f}} < \tau$. Following the calculations in Ref.~ \cite{Liou:2013qya}, one finds
 \begin{align}
\label{DLAexp2}
\delta\lan p_\perp^2(L)\ran 
\,=\,% \frac{2\alpha_s C_F}{\pi} 
\abar\int_{\tau_0}^L \rmd \tau\,\hat q_0(\tau) \int_{\lambda(\tau)}^{\tau}
\frac{\rmd \tau_{\rm f}}{\tau_{\rm f}}
  \int^{\mathcal{Q}_0^2(L)\tau_{\rm f}}_{\hat q_0(\tau)\tau_{\rm f}^2} \frac{\rmd \omega}{\omega} \,.
  \end{align}
 The limits in the $\omega$-integration can be understood as follows. The lower limit comes from the fact that the function $D$ in \eqn{Dcapp} vanishes exponentially when $|\Omega(\tau) \tau_{\text{f}}|\gtrsim 1$. The upper limit represents the validity limit for the small-$r^2$ expansion in \eqn{equ:Sexp}: this expansion assumes $r^2 \ll  \tau_{\rm f}/\omega$, where $r^2\sim 1/\mathcal{Q}_0^2(L)$. Finally,  the lower limit $\lambda$ on the formation time $\tau_{\rm f}$ follows from a careful analysis of the kinematics of the in-medium scattering  \cite{Liou:2013qya}.
After the change of integration variable $\omega \to k_\perp^2\equiv \omega/\tau_{\rm f}$, \eqn{DLAexp2} becomes identical with \eqref{DLAexp} that was employed in the main text.

\providecommand{\href}[2]{#2}\begingroup\raggedright\endgroup
\end{document}